\documentclass[pra,preprint,showpacs,preprintnumbers,amsmath,amssymb,superscriptaddress,floatfix,showkeys]{revtex4}
\usepackage{amsfonts}
\usepackage{graphicx}
\usepackage{dcolumn}
\usepackage{bm}
\usepackage{amsmath}
\usepackage{amssymb}
\usepackage{longtable}
\begin{document}

%
%
%
\title{Relativistic Calculation Of Two-Electron One-Photon And
  Hypersatellite Transition  Energies For $12\leq Z\leq30$ Elements}
%
%
%
%
\author{M. C. Martins}
 \email{mdmartins@fc.ul.pt} 
 \affiliation{Centro de F{\'\i}sica At{\'o}mica da Universidade de Lisboa, \\
  Av. Prof. Gama Pinto 2, 1649-003 Lisboa, Portugal}
 \affiliation{Departamento F{\'\i}sica, Faculdade de Ci{\^e}ncias,
 Universidade de Lisboa, Campo Grande, 1749-016 Lisboa, Portugal}
\author{A. M. Costa}
 \affiliation{Centro de F{\'\i}sica At{\'o}mica da Universidade de Lisboa, \\
  Av. Prof. Gama Pinto 2, 1649-003 Lisboa, Portugal}
 \affiliation{Departamento F{\'\i}sica, Faculdade de Ci{\^e}ncias,
 Universidade de Lisboa, Campo Grande, 1749-016 Lisboa, Portugal}
\author{J. P. Santos} 
 \affiliation{Centro de F{\'\i}sica At{\'o}mica da Universidade de Lisboa, \\
  Av. Prof. Gama Pinto 2, 1649-003 Lisboa, Portugal}
 \affiliation{Departamento de F{\'\i}sica, Faculdade de Ci{\^e}ncias e Tecnologia, \\
  Universidade Nova de Lisboa, Monte de Caparica, 2825-114 Caparica, Portugal}
\author{F. Parente}
 \affiliation{Centro de F{\'\i}sica At{\'o}mica da Universidade de Lisboa, \\
  Av. Prof. Gama Pinto 2, 1649-003 Lisboa, Portugal}
 \affiliation{Departamento F{\'\i}sica, Faculdade de Ci{\^e}ncias,
 Universidade de Lisboa, Campo Grande, 1749-016 Lisboa, Portugal}
\author{P. Indelicato}
 \affiliation{Laboratoire Kastler-Brossel,
 {\'E}cole Normale Sup{\'e}rieure et Universit{\'e} Pierre et Marie 
 Curie, Case 74, 4 place Jussieu,
 F-75252 Paris CEDEX 05, France}

\date{\today}
%

%
%
\begin{abstract}
  Energies of two-electron one-photon transitions from initial double
  K-hole states were computed using the Dirac-Fock model. The
  transition energies of competing processes, the K$\alpha$
  hypersatellites, were also computed. The results are compared to
  experiment and to other theoretical calculations.
\end{abstract}

%
%


\maketitle

%

\section{Introduction}

Energies and transition rates for some radiative processes in atoms
initially bearing two K-shell holes (two-electron one-photon and
one-electron one-photon transitions) were evaluated in this work.
This kind of atom, in which an entire inner shell is empty while the
outer shells are occupied, was first named hollow by Briand
~\textit{et al.}~\cite{888}. Hollow atoms are of great importance for
studies of ultrafast dynamics in atoms far from equilibrium and have
possible wide-ranging applications in physics, chemistry, biology, and
materials science~\cite{889}.

\begin{figure}
[hh]
\begin{center}
\includegraphics[
height=2.4976in,
width=3.7559in
]%
{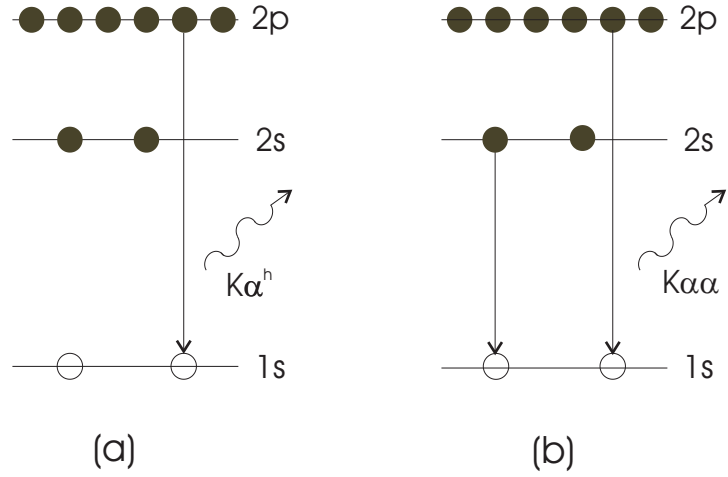}
\caption{(a) One-electron one-photon transition that produces the
K$\alpha^{\text{h}}$ hypersatellite lines; (b) two-electron one-photon
transition that produces the K$\alpha\alpha$ lines.}%
\label{fig02}%
\end{center}
\end{figure}

A mono-ionized atom with a K-shell vacancy can decay through an L
$\rightarrow$ K electron transition with the emission of x-ray
radiation called, in the Siegbahn notation, the K$\alpha$ diagram
line. A one-electron transition line for which the initial state has
two vacancies in the same shell is called a hypersatellite line. This
is the case when a double ionized K-shell state decays through the
transition of one L-shell electron (Fig.~\ref{fig02}-a),
K$^{-2}\rightarrow$ K$^{-1}$L$^{-1}$, which is denoted by
K$\alpha^{\text{h}}$.
A competing, less probable, process of radiative de-excitation from
this state is the simultaneous transition of two correlated electrons
from higher shells, the K$^{-2}\rightarrow$ L$^{-2}$ transitions,
accompanied by the emission of a single photon carrying the total
energy, called K$\alpha\alpha$ (Fig.~\ref{fig02}-b).
Predictions of this decaying process can be found in early papers at
the beginning of the twentieth century by Heisenberg~\cite{885} and
Condon~\cite{886}, but it has only been observed since
1975~\cite{787,734}.

For comparison with theory,
we should distinguish between experiments in which the initial atomic
excitation uses electrons as projectiles ~\cite{783,785,784},
photoionization or nuclear decay ~\cite{786}, and experiments using
heavy ions ~\cite{735,787,734}. In the latter case, the probability of
multiple ionization is usually very high, leading to unreliable
determination of energy values.
One of the reasons for the scarcity of accurate experimental data stems
from the very low intrinsic probability of creating a state with just two
K-shell holes.

Theoretical calculations so far have mainly used perturbation theory
and were performed in a non-relativistic approach ~\cite{880,896} with
the exception of Chen et al. work ~\cite{308}, in which a
Dirac-Hartree-Slater approach was used.
Indeed, for medium-$Z$ atoms the K-shell electrons are already
significantly relativistic, thus calling for relativistic methods in
atomic data calculations.
In this work we used the Dirac-Fock model to compute transition
energies for two-electron one-photon transitions arising from the
de-excitation of double 1s hole states leading to final states with
two L-shell holes in atoms with $12\leq Z\leq30$. This is the region
of atomic numbers where the transition from the LS to the intermediate
coupling scheme occurs. This transition is reflected directly in the
K$\alpha^{\text{h }}$ lines relative intensities. We also computed
the energies of the competing hypersatellite transitions. The results
are compared to experiment and other theoretical calculations.

%
%
\section{Calculation of atomic wave functions and energies}

Bound state wave functions and radiative transition probabilities were
calculated using the multi-configuration Dirac-Fock program of J. P.
Desclaux and P. Indelicato ~\cite{32,62}.
The program was used in single-configuration mode because correlation
was found to be unimportant. The wave functions of the initial and
final states were computed independently, that is, atomic orbitals
were fully relaxed in the calculation of the wave function for each
state, and non-orthogonality was taken in account in transition
probabilities calculations.

In order to obtain a correct relationship between many-body methods and 
quantum electrodynamics (QED)~\cite{47,231,229,230}, one should start 
from the no-pair Hamiltonian
\begin{equation}
\label{eq:hamilnopai}
        {\cal H}^{\mbox{{\tiny no pair}}}=\sum_{i=1}^{N}{\cal
        H}_{D}(r_{i})+\sum_{i<j}{\cal V}(|\bm{r}_{i}-\bm{r}_{j}|),
\end{equation}
where ${\cal H}_{D}$ is the one electron Dirac operator and ${\cal V}$
is an operator representing the electron-electron interaction of order
one in $\alpha$, properly set up between projection operators
$\mathit{\Lambda}_{ij}^{++}=\mathit{\Lambda}_{i}^{+}\mathit{\Lambda}_{j}^{+}$
to avoid coupling positive and negative energy states
\begin{equation}
\mathcal{V}_{ij}=\mathit{\Lambda}_{ij}^{++}V_{ij}\mathit{\Lambda}_{ij}^{++}.
\end{equation}
The expression of $V_{ij}$ in the Coulomb gauge and in atomic units is
%
\begin{subequations}
\label{eq:eeinter}
\begin{align}
         V_{ij} =& \,\,\,\, \frac{1}{r_{ij}} \label{eq:coulop} \\
         &-\frac{\bm{\alpha}_{i} \cdot \bm{\alpha}_{j}}{r_{ij}} 
\label{eq:magop} \\ 
         & - \frac{\bm{\alpha}_{i} \cdot
         \bm{\alpha}_{j}}{r_{ij}} 
[\cos\left(\frac{\omega_{ij}r_{ij}}{c}\right)-1]
         \nonumber \\
        & + c^2(\bm{\alpha}_{i} \cdot
         \bm{\nabla}_{i}) (\bm{\alpha}_{j} \cdot
         \bm{\nabla}_{j})
         \frac{\cos\left(\frac{\omega_{ij}r_{ij}}{c}\right)-1}{\omega_{ij}^{2} 
r_{ij}},
         \label{eq:allbreit}
\end{align}
\end{subequations}
%
where $r_{ij}=\left|\bm{r}_{i}-\bm{r}_{j}\right|$ is the
inter-electronic distance, $\omega_{ij}$ is the energy of the photon
exchanged between the two electrons, $\bm{\alpha}_{i}$ are the Dirac
matrices and $c=1/\alpha$ is the speed of light, $\alpha$ being the
fine structure constant. We use the Coulomb gauge as it has been
demonstrated that it provides energies free from spurious
contributions at the ladder approximation level and must be used in
many-body atomic structure calculations~\cite{63,238}.

The term (\ref{eq:coulop}) represents the Coulomb interaction, the
term (\ref{eq:magop}) is the Gaunt (magnetic) interaction, and the
last two terms (\ref{eq:allbreit}) stand for the retardation operator.
In this expression the $\bm{\nabla}$ operators act only on $r_{ij}$
and not on the following wave functions.

By a series expansion of the operators in expressions~(\ref{eq:magop})
and (\ref{eq:allbreit}) in powers of $\omega_{ij}r_{ij}/c \ll 1$ one
obtains the Breit interaction, which includes the leading retardation
contribution of order $1/c^{2}$. The Breit interaction is, then, the
sum of the Gaunt interaction (\ref{eq:magop}) and the Breit
retardation
\begin{equation}
\label{eq:breit}
B^{\text{\scriptsize{R}}}_{ij} =
{\frac{\bm{\alpha}_i\cdot\bm{\alpha}_j}{2r_{ij}}} - 
\frac{\left(\bm{\alpha}_i\cdot\bm{r}_{ij}\right)\left(\bm{\alpha}_j
\cdot\bm{r}_{ij}\right)}{{2r_{ij}^3}}.
\end{equation}
In the many-body part of the calculation the electron-electron
interaction is described by the sum of the Coulomb and the Breit
interactions. Higher orders in $1/c$, deriving from the difference
between Eqs.~(\ref{eq:allbreit}) and (\ref{eq:breit}) are treated here
only as a first order perturbation.

%
All calculations are done for finite nuclei using uniformly charged
spheres. 

Finally, from a full QED treatment, one also obtains the radiative
corrections (important for the innermost shells) to the
electron-nucleus interaction (self-energy and vacuum polarization).
The one-electron self-energy is evaluated using the one-electron
values of Mohr and coworkers ~\cite{115,114,116}.  The self-energy
screening is treated with the Welton method developed in
Refs.~\cite{58,56,53,847}.  This method yields results in close
agreement (better than 5\%) with \textit{ab initio} methods based on
QED ~\cite{242,263,288}, without the huge amount of effort involved.
The vacuum polarization is evaluated as described in Ref.~\cite{914}.
The Uelhing contribution is evaluated to all orders by being included
in the self-consistent field (SCF). The Wichmann and Kroll and
K\"all\'en and Sabry contributions are included perturbatively. All
three contributions are evaluated using the numerical procedure from
Refs.~\cite{158,277}.

Breit and QED contributions to the energy of some levels are shown in
Table~\ref{tab_contrib} for Mg, Ca and Zn.

%
%
\section{Results}

We calculated the energies of the K$\alpha^{\text{h }}$ hypersatellite
transitions and the K$\alpha\alpha$ two-electron one-photon
transitions for atoms with $12\leq Z\leq30$. 
%

Depending on the configurations of the initial and final states, for
the different values of $Z$, the number of transitions that must be
dealt with may range from only two, when the initial state has only
closed shells, to several hundred, when unfilled shells exist.

For Mg, Ar, Ca and Zn the 1s$^{-2}$ ground configuration corresponds
to only one level, the $^{1}$S$_{0}$ level, and each of the K$\alpha \alpha $ or K$%
\alpha ^{h}$ lines is identified by a precise level transition,

\bigskip 
\begin{tabular}{ll}
K$\alpha _{2}\alpha _{3}$: &  $1s^{-2\text{ \ }1}$S$_{0}\rightarrow
2s^{-1}2p^{-1}$ $^{1}$P$_{1}$ \\ 
K$\alpha _{1}\alpha _{3}$: $\ $ & $1s^{-2\text{ \ }1}$S$_{0}\rightarrow
2s^{-1}2p^{-1}$ $^{3}$P$_{1}$ \\ 
K$\alpha _{2}^{h}$:  & $1s^{-2\text{ \ }1}$S$_{0}\rightarrow 1s^{-1}2p^{-1}$ 
$^{1}$P$_{1}$ \\ 
K$\alpha _{1}^{h}$: $\ $ & $1s^{-2\text{ \ }1}$S$_{0}\rightarrow
1s^{-1}2p^{-1}$ $^{3}$P$_{1}$%
\end{tabular}
\bigskip

To be able to compare our theoretical transition energy values with
experiment and non-relativistic calculations by other authors, we must
define the statistical average energy of a line. The energy of all
individual $i\rightarrow f$ transitions in the $X$ line, from an
initial level $i$, $E_{if}$, weighted by the corresponding transition
probability, $W_{if}$, yields the average energy of the $X$ line
coming from level $i$, $E_{X}\left( i\right) $:
\begin{equation}
E_{X}\left(i\right) = \dfrac{\sum\limits_{f\left(  X\right)  }E_{if}W_{if}}{\sum
\limits_{f\left(  X\right)  }W_{if}}
\label{eq0031}
\end{equation}
In Eq. (\ref{eq0031}), $f\left( X\right) $ runs over all possible
final levels in the radiative de-excitation leading to the $X$ line,
from a specific initial level $i$.
Assuming that all states of a $\gamma$ configuration are equally
populated, the resulting $E_{X}\left( i\right) $ energies were then
weighted by the statistical weight of level $i$, $g\left( i\right) $,
leading to the statistical average energy $E_{X}^{\text{SA}}$ for the
$X$ line:
\begin{equation}
  E_{X}^{\text{SA}} =\dfrac{1}{g\left( \gamma\right)
    }\sum_{i}g\left( i\right) E_{X}\left(i\right).
\label{eq003}
\end{equation}
Here, ${g\left( \gamma\right)}$ is the statistical weight of the
$\gamma$ configuration. We estimate the uncertainty of transition
energy values has being of the order of 1 eV.

\begin{figure}
  [h]
\begin{center}
\includegraphics[
height=3.3961in,
width=3.813in
]%
{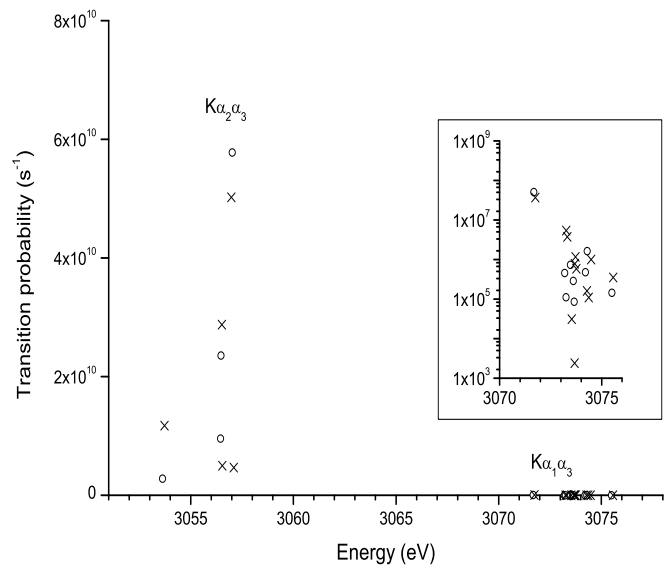}
\caption{Transition probabilities for all K$\alpha\alpha$ two-electron
one-photon transitions in aluminium. The transitions for the
initial levels $^2$P$_{1/2}$ and $^2$P$_{3/2}$ are represented,
respectively, by $\circ$ and $\times$.}%
\label{fig04a}%
\end{center}
\end{figure}

In Table \ref{tab:al}, we tabulate the results obtained in this work
for the two-electron one-photon radiative transition energies and
probabilities from a double K-hole state in aluminium.
In this table, the transitions in the K$\alpha\alpha$ lines were
ordered by energy: two groups of transitions well separated in energy
are clearly identified, which we interpret as being the
K$\alpha_{2}\alpha_{3}$ and K$\alpha_{1}\alpha_{3}$ lines,
respectively. Further details can be found in~\cite{919}. Afterwards,
the transitions were grouped, within each of these two groups of
transitions, by their initial level, either $^{2}$P$_{1/2}$ or
$^{2}$P$_{3/2}$.
Transition probabilities for all K$\alpha\alpha$ two-electron
one-photon transitions in aluminium are shown in Fig.~\ref{fig04a}.

As can be seen in Table~\ref{tab:al}, for aluminium the values of the
average energy $E_{X}$ for transitions starting from different initial
levels are very similar.  This is also evident in Table~\ref{tab:ti},
where the average energy values of the x-ray lines for different
initial levels of titanium are presented.  It is worth mentioning that
401 transition energies were computed to obtain the results
corresponding to the K$\alpha\alpha$ lines of the latter case.

To avoid time-consuming calculations, some authors have calculated
just the 2s$^{2}$2p$^{6}\rightarrow$ 1s$^{2}$2s2p$^{5}$ transition
energy for all values of $Z$, thus neglecting the interaction with the
outer electrons. To check the magnitude of the error arising from this
simplification, we calculated the transition energies for the atoms
with $Z=12,13,18,20,21,22,28,30$, first taking in account all
electrons and, in a separate calculation, including in the initial
configuration only electrons in the L-shell.

\begin{figure}
  [h]
\begin{center}
\includegraphics[
height=3.205in,
width=3.3364in
]%
{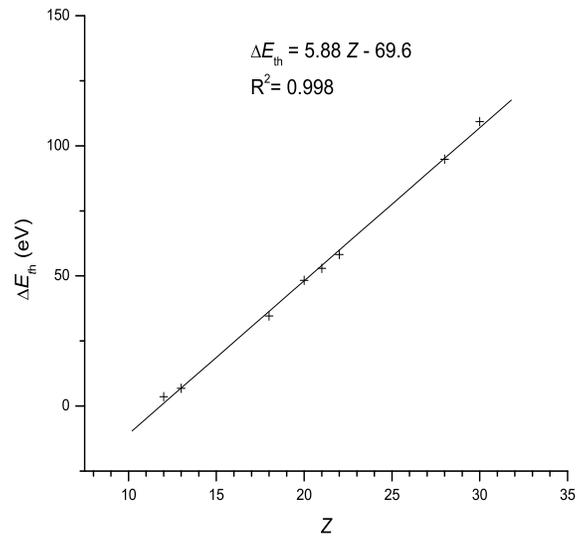}%
\caption{The differences $\Delta E_{\text{th}}$, as a function of
  atomic number $Z$, between the transition energies calculated for
  the K$\alpha_{2}\alpha_{3}$ line, using only the 2s$^{2}$ 2p$^{6}$
  configuration and using the 1s$^{-2}$ configuration (+). The
  straight line represents the linear regression to these energy
  differences.}
\label{fig05}%
\end{center}
\end{figure}

For example, in the case of aluminium K$\alpha \alpha$ lines, we
calculated the transition energy for both the
2s$^{2}$2p$^{6}$3s$^{2}$3p$\,\rightarrow$1s$^{2}$2s 2p$^{5}$3s$^{2}$3p
and the 2s$^{2}$2p$^{6}\,\rightarrow$1s$^{2}$2s 2p$^{5}$ transitions.
In the particular case of K$\alpha_{2}\alpha_{3}$, we found a 6.8 eV
energy difference between the two energy values, out of the 3056.54 eV
transition energy (a difference of 0.2 \%).
Table~\ref{tab:energies} gives the energy values of the
K$\alpha_{2}\alpha_{3}$ line for the 8 elements considered, obtained
through the two approaches described.
%
%
The energy differences, $\Delta E_{\text{th}}$, thus obtained were
then fitted to a straight line as a function of $Z$
(Fig.~\ref{fig05}). We found that the fit is quite good, presenting a
correlation coefficient of 0.998.  Using the results of this fitting
process, and calculations where only the 2s$^{2}$2p$^{6}$ electrons
were included in the initial states, we obtained
K$\alpha_{2}\alpha_{3}$ transition energies for the atoms with the
remaining $Z$ values.

Similar behaviour is also observed from the other lines obtained in the
decay of a double K-hole state (K$\alpha_{1}\alpha_{3}$,
K$\alpha_{2}^{\text{h}}$, K$\alpha_{1}^{\text{h}}$). Thus, using the
same method, we were able to obtain the transition energy values for
other atoms, with values of $Z$ between 12 and 30, for which complete
calculations would involve time-consuming work. For example, for iron
($Z=26$) it would be necessary to calculate around five thousand
transition energy values to obtain the energy of the K$\alpha
_{2}^{\text{h}}$ and K$\alpha_{1}^{\text{h}}$ hypersatellite or
K$\alpha _{2}\alpha_{3}$ and K$\alpha _{1}\alpha _{3}$ lines.

The results, for all elements with $12\leq Z\leq30$, are presented in
Table ~\ref{tab:all_z_kah} and Table ~\ref{tab:all_z_kaa}, where they
are compared with other theoretical calculations and experimental
values.  A comparison between the results of this work for
K$\alpha_{2}^{\text{h}}$ line energies and other available results is
presented in Fig.~\ref{fig06}.

%
%
\section{Discussion and conclusions}

In this work, we computed the energy of K$\alpha_{2}^{\text{h}}$,
K$\alpha _{1}^{\text{h}}$, K$\alpha_{2}\alpha_{3}$ and
K$\alpha_{1}\alpha_{3}$ lines in the framework of the Dirac-Fock
approximation for elements with atomic number $12\leq Z\leq30$. For
selected elements we performed two different calculations: first we
took $1$s$^{-2}$ as the initial configuration, then we repeated the
calculation, considering the 2s$^{2}$2p$^{6}$ configuration as the
initial one. We fitted the differences between the values obtained in
the two calculations, as a function of $Z$, to a straight line. We
used this result to make a correction to the energies of the remaining
elements calculated using the 2s$^{2}$2p$^{6}$ configuration as the
initial one.

These results can be compared with experimental work in which K-holes
were obtained using electron bombardment, photo ionization or
radioactive decay. Other methods of producing K shell holes, like ion
bombardment, will inevitably produce extra holes in the atom, leading
to shifts in the measured value of the transition energy, unless the
resolution obtained with the detection process is high enough to allow
for separation of the peaks resulting from multiple vacancies.

\begin{figure}
[h]
\begin{center}
\includegraphics[
height=9.40cm,
width=12.26cm
]%
{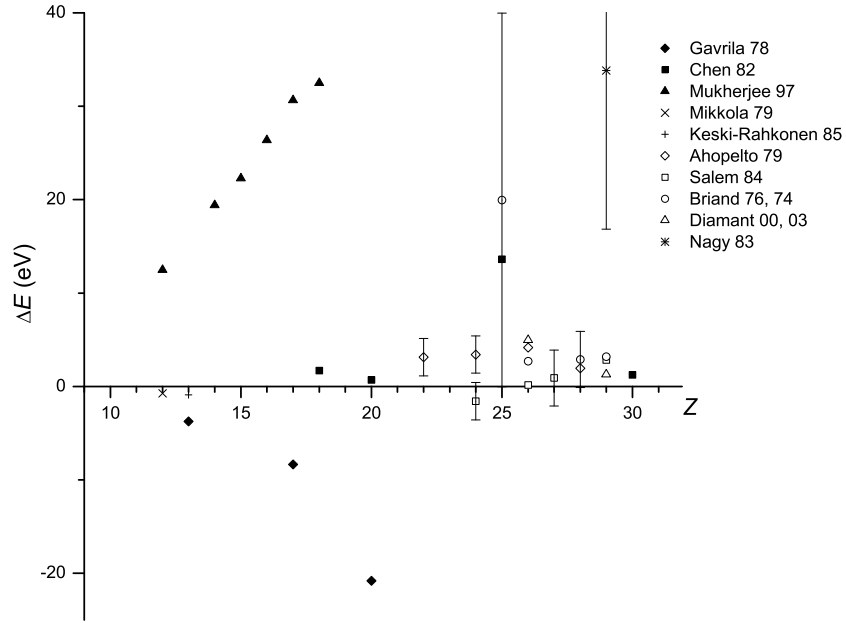}%
\caption{Comparison between available calculated and measured values of
  K$\alpha_{2}^{\text{h}}$ energy and those of the present work, for
  $12\leq Z\leq 30$. $\Delta E$ stands for the difference between
  other energy values and the values obtained in this work.}
\label{fig06}%
\end{center}
\end{figure}

The values obtained in this work for the K$\alpha_{2}^{\text{h}}$ and
K$\alpha_{1}^{\text{h}}$ line energies are in excellent agreement with
the experimental results (see Fig.~\ref{fig06} for
K$\alpha_{2}^{\text{h}}$ energies).  We particularly emphasize the
agreement with the measured K$\alpha_{2}^{\text{h}}$ energy value of
Mikkola et al.~\cite{897}, for $Z=12$, Keski-Rahkonen et
al.~\cite{913} for $Z=13$, , and Diamant et al.~\cite{890}, for
$Z=29$, due to the reported high experimental precision.  These
authors resolved the x-ray lines using crystal spectrometers.

In Table ~\ref{tab:all_z_kah} we present a comparison between our
calculated values and the available experimental and theoretical data,
in particular those obtained by Chen et al.~\cite{308} in the
framework of the Dirac-Hartree-Slater (DHS) method for $Z=18,20,25$
and $30$.  The differences between the present calculations and the
DHS calculations are probably due mainly to the differences in wave
functions (ours were obtained with the optimized level (OL) method and
the DHS ones with the configuration-average level method).
Additionally, we used an electron-electron operator that avoids
coupling between positive and negative states, and we included higher
orders in $1/c$ of the Breit interaction.

Our results are closer to experiment than other theoretical results
for the K$\alpha_{2}\alpha_{3}$ line; no previous calculation has been
reported, to our knowledge, for the K$\alpha_{1}\alpha_{3}$ line. A
more detailed comparison with the available experimental results
reveals that, for $24\leq Z\leq29$, our values lie systematically
higher (between 0.2\% and 0.4\%) than the experimental values, outside
the reported experimental uncertainty in all cases ~\cite{784,786}.
Regarding the value of Auerhammer et al.~\cite{783} for the Al
($Z=13$) K$\alpha_{2}\alpha_{3}$ line, some comments are in order.
These authors report the K$^{-2}\rightarrow$ L$^{-2}$ two-electron
one-photon lines as well as the K$^{-2}$L$^{-n}\rightarrow$
L$^{-\left( 2+n\right) }$ satellite lines, with $n=1,\ldots,4$.
They used the calculated values of Tannis et al.~\cite{916} to propose
the identification of most of their measured transition lines. The
line with energy difference $\Delta E=E\left(
  \text{K}\alpha\alpha\right) -2E\left(
  \text{K}\alpha_{1}\right)=103.4$ eV was labelled by them as the
K$\alpha_{2}\alpha_{3}$ line. We suggest that this label should be
attributed instead to their unidentified line with energy $\Delta
E=84.8$ eV, in much closer agreement with our calculated value $\Delta
E=82.2$ eV.



%
%
%
\begin{acknowledgments}

  This research was supported in part by FCT project
  POCTI/FAT/50356/2002 financed by the European Community Fund FEDER.
\end{acknowledgments}

%
%
%
\bibliography{jps}

%
%
%

\newpage

\newpage


\begin{table}                                               
\begin{center}                                              
\caption{QED and Breit contributions to the total energy, $E$, in eV, for Mg,
  Ca and Zn.}
\label{tab_contrib}                                              
\scriptsize 
\begin{tabular}{lccccccccccc}
\hline\hline
&  &  &  &  &  &  &  &  &  \\ 
& \multicolumn{3}{c}{Mg ($Z$=12)} && \multicolumn{3}{c}{Ca ($Z$=20)} & &\multicolumn{3}{c}{Zn ($Z$=30)} \\ 
\cline{2-4}\cline{6-8}\cline{10-12}
& 1s$^{-2}$ & 1s$^{-1}$2p$^{-1}$ & 2s$^{-1}$2p$^{-1}$ & &1s$^{-2}$ & 1s$^{-1}$2p$^{-1}$ & 2s$^{-1}$2p$^{-1}$ && 1s$^{-2}$ & 1s$^{-1}$2p$^{-1}$ & 2s$^{-1}$2p$^{-1}$ \\ 
& $^{1}$S$_{0}$ & $^{3}$P$_{1}$ & $^{3}$P$_{1}$ && $^{1}$S$_{0}$ & $^{3}$P$_{1}$ & $^{3}$P$_{1}$ && $^{1}$S$_{0}$ & $^{3}$P$_{1}$ & $^{3}$P$_{1}$ \\ 
\hline
&  &  &  &  &  &  &  &  &  \\ 
$E$ & \multicolumn{1}{r}{-2663.734} & \multicolumn{1}{r}{-4038.075} & \multicolumn{1}{r}{-5264.540} && \multicolumn{1}{r}{-10117.187} & \multicolumn{1}{r}{-14013.573} & \multicolumn{1}{r}{-17642.276} && \multicolumn{1}{r}{-29042.059} & \multicolumn{1}{r}{-37997.033} & \multicolumn{1}{r}{-46466.031} \\ 
QED & \multicolumn{1}{r}{0.057} & \multicolumn{1}{r}{0.327} & \multicolumn{1}{r}{0.549}& & \multicolumn{1}{r}{0.420} & \multicolumn{1}{r}{1.989} & \multicolumn{1}{r}{3.311} && \multicolumn{1}{r}{1.878} & \multicolumn{1}{r}{8.015} & \multicolumn{1}{r}{13.235} \\ 
Breit & \multicolumn{1}{r}{0.139} & \multicolumn{1}{r}{0.249} & \multicolumn{1}{r}{0.820} && \multicolumn{1}{r}{1.010} & \multicolumn{1}{r}{1.822} & \multicolumn{1}{r}{4.748} && \multicolumn{1}{r}{4.886} & \multicolumn{1}{r}{8.158} & \multicolumn{1}{r}{18.454} \\ 
&  &  &  &  &  &  &  &  &  \\ 
\hline\hline
\end{tabular}
\end{center}                                     
\end{table}                                      


%
%
%

\begin{table}                                               
\begin{center}                                              
\caption{Two-electron one-photon radiative transition energies ($E_{if}$) and
  probabilities ($W_{if}$) for the 2s$^{2}$ 2p$^{6}$ 3s$^{2}$
  3p$\rightarrow $1s$^{2}$ 2s 2p$^{5}$ 3s$^{2}$ 3p transition in
  aluminium. $E_{X}$ is defined by Eq.~(\ref{eq0031}) and $E_{X}^{SA}$
  is defined by Eq.~(\ref{eq003}). Superscripts $^{\#i}$ are added to
  distinguish identical terms in the same configuration, where
  $^{\#1}$ stands for the term with the lowest energy. The transtion
  energy values are in eV.}
\label{tab:al}                                              
\scriptsize 
\begin{tabular}{cccclcc}
\hline\hline
\\
& Initial Level & Final Level  & $E_{if}$ & $W_{if}$ (s$^{-1}$)  & $E_{X}$
 & $E_{X}^{SA}$  \\ 
\\
\hline
\\
K$\alpha _{2}\alpha _{3}$ & $^{2}P_{1/2}$ & $^{2}S_{1/2}$$^{\#2}$ & 3053.62 & 2.79$\times
10^{9}$ &  &  \\ 
&  & $^{2}P_{3/2}$$^{\#2}$ & 3056.45 & 9.53$\times 10^{9}$ &  &  \\ 
&  & $^{2}P_{1/2}$$^{\#2}$ & 3056.47 & 2.36$\times 10^{10}$ &  &  \\ 
&  & $^{2}D_{3/2}$$^{\#2}$ & 3057.02 & 5.78$\times 10^{10}$ & 3056.72 &  \\ 
& $^{2}P_{3/2}$ & $^{2}S_{1/2}$$^{\#2}$ & 3053.72 & 1.17$\times 10^{10}$ &  &  \\ 
&  & $^{2}P_{3/2}$$^{\#2}$ & 3056.52 & 2.87$\times 10^{10}$ &  &  \\ 
&  & $^{2}P_{1/2}$$^{\#2}$ & 3056.54 & 4.97$\times 10^{9}$ &  &  \\ 
&  & $^{2}D_{5/2}$$^{\#2}$ & 3056.97 & 5.02$\times 10^{10}$ &  &  \\ 
&  & $^{2}D_{3/2}$$^{\#2}$ & 3057.09 & 4.69$\times 10^{9}$ & 3056.45 & 3056.54 \\  
\\
K$\alpha _{1}\alpha _{3}$ & $^{2}P_{1/2}$ & $^{2}S_{1/2}$$^{\#1}$ & 3071.68 & 5.06$\times
10^{7}$ &  &  \\ 
&  & $^{2}P_{1/2}$$^{\#1}$ & 3073.19 & 4.54$\times 10^{5}$ &  &  \\ 
&  & $^{2}P_{3/2}$$^{\#1}$ & 3073.25 & 1.11$\times 10^{5}$ &  &  \\ 
&  & $^{2}D_{3/2}$$^{\#1}$ & 3073.47 & 7.41$\times 10^{5}$ &  &  \\ 
&  & $^{4}P_{1/2}$ & 3073.61 & 2.87$\times 10^{5}$ &  &  \\ 
&  & $^{4}P_{3/2}$ & 3073.65 & 8.59$\times 10^{4}$ &  &  \\ 
&  & $^{4}D_{1/2}$ & 3074.21 & 4.77$\times 10^{5}$ &  &  \\ 
&  & $^{4}D_{3/2}$ & 3074.29 & 1.64$\times 10^{6}$ &  &  \\ 
&  & $^{4}S_{3/2}$ & 3075.50 & 1.45$\times 10^{5}$ & 3071.85 &  \\ 
& $^{2}P_{3/2}$ & $^{2}S_{1/2}$$^{\#1}$ & 3071.75 & 3.71$\times 10^{7}$ &  &  \\ 
&  & $^{2}P_{1/2}$$^{\#1}$ & 3073.26 & 5.44$\times 10^{6}$ &  &  \\ 
&  & $^{2}P_{3/2}$$^{\#1}$ & 3073.32 & 3.78$\times 10^{6}$ &  &  \\ 
&  & $^{2}D_{3/2}$$^{\#1}$ & 3073.54 & 5.61$\times 10^{4}$ &  &  \\ 
&  & $^{4}P_{1/2}$ & 3073.68 & 6.84$\times 10^{3}$ &  &  \\ 
&  & $^{2}D_{5/2}$$^{\#1}$ & 3073.69 & 7.90$\times 10^{5}$ &  &  \\ 
&  & $^{4}P_{3/2}$ & 3073.72 & 1.17$\times 10^{6}$ &  &  \\ 
&  & $^{4}P_{5/2}$ & 3073.76 & 6.48$\times 10^{5}$ &  &  \\ 
&  & $^{4}D_{1/2}$ & 3074.28 & 1.63$\times 10^{5}$ &  &  \\ 
&  & $^{4}D_{3/2}$ & 3074.36 & 1.12$\times 10^{5}$ &  &  \\ 
&  & $^{4}D_{5/2}$ & 3074.48 & 9.99$\times 10^{5}$ &  &  \\ 
&  & $^{4}S_{1/2}$ & 3075.57 & 3.68$\times 10^{5}$ & 3072.23 & 3072.10 \\ 
\\
\hline\hline
\end{tabular}                                       
\end{center}                                     
\end{table}                                      




\begin{table}                                               
\begin{center}                                              
\caption{Titanium two-electron one-photon radiative transition average
  energies in eV of the $X$ line coming from the initial level.
  $E_{X}^{SA}$ is defined by Eq. (\ref{eq003}).}
\label{tab:ti}                                              
\scriptsize 
\begin{tabular}{cccc}
\hline\hline
&  &  &  \\ 
& Initial Level & $E_{X}$  & $E_{X}^{SA}$  \\ 
&  &  &  \\ \hline
&  &  &  \\ 
K$\alpha _{2}\alpha _{3}$ & $^3P_0$ & 9144.96 &  \\ 
& $^1S_0$ & 9144.93 &  \\ 
& $^3P_1$ & 9144.99 &  \\ 
& $^3F_2$ & 9144.97 &  \\ 
& $^1D_2$ & 9144.97 &  \\ 
& $^3P_2$ & 9145.05 &  \\ 
& $^3F_3$ & 9145.04 &  \\ 
& $^3F_4$ & 9145.12 &  \\ 
& $^1G_4$ & 9145.05 & 9145.04 \\ 
&  &  &  \\ 
K$\alpha _{1}\alpha _{3}$ & $^3P_0$ & 9174.22 &  \\ 
& $^1S_0$ & 9175.64 &  \\ 
& $^3P_1$ & 9174.42 &  \\ 
& $^3F_2$ & 9173.64 &  \\ 
& $^1D_2$ & 9173.98 &  \\ 
& $^3P_2$ & 9175.19 &  \\ 
& $^3F_3$ & 9174.20 &  \\ 
& $^3F_4$ & 9175.37 &  \\ 
& $^1G_4$ & 9175.29 & 9174.72 \\ 
&  &  &  \\ 
\hline\hline
\end{tabular}                                       
\end{center}                                     
\end{table}                                      




\begin{table}                                               
\begin{center}                                              
\caption{Transition energies in eV for K$\alpha_{2}\alpha_{3}$ calculated
  using all electrons ($E_{X}^{\text{SA}}$) and using only the
  2s$^{2}$ 2p$^{6}$ electrons ($E$*), respectively. $\Delta E_{\text{th}}$
  indicates the difference between these energy values: $\Delta
  E_{\text{th}}$=$E$*-$E_{X}^{\text{SA}}$.}
\label{tab:energies}                                              
\scriptsize 
\begin{tabular}{clccc}
\hline\hline
$Z$ & Initial Configuration & $E_{X}^{\text{SA}}$  & $E$*  & $\Delta E_{\text{th}}$  \\ 
\hline
&  &  &  &  \\ 
12 & 2s$^{2}$ 2p$^{6}$ 3s$^{2}$ & 2585.45 & 2589.01 & 3.56 \\ 
13 & 2s$^{2}$ 2p$^{6}$ 3s$^{2}$ 3p & 3056.54 & 3063.37 & 6.83 \\ 
18 & 2s$^{2}$ 2p$^{6}$ 3s$^{2}$ 3p$^{6}$ & 6022.29 & 6056.94 & 34.65 \\ 
20 & 2s$^{2}$ 2p$^{6}$ 3s$^{2}$ 3p$^{6}$ 4s$^{2}$ & 7497.79 & 7546.10 & 48.31\\ 
21 & 2s$^{2}$ 2p$^{6}$ 3s$^{2}$ 3p$^{6}$ 3d 4s$^{2}$ & 8300.74 & 8353.62 & 52.88 \\ 
22 & 2s$^{2}$ 2p$^{6}$ 3s$^{2}$ 3p$^{6}$ 3d$^{2}$ 4s$^{2}$ & 9145.04 & 9203.25 & 58.21 \\ 
28 & 2s$^{2}$ 2p$^{6}$ 3s$^{2}$ 3p$^{6}$ 3d$^{8}$ 4s$^{2}$ & 15098.34 & 15193.17 & 94.83 \\ 
30 & 2s$^{2}$ 2p$^{6}$ 3s$^{2}$ 3p$^{6}$ 3d$^{10}$ 4s$^{2}$ & 17423.97 & 17533.31 & 109.34 \\ 
&  &  &  &  \\ 
\hline\hline
\end{tabular}                                       
\end{center}                                     
\end{table}                                      




\begin{table}                                               
\begin{center}                                              
\caption{K$\alpha _{2}^{\text{h}}$ and K$\alpha_{1}^{\text{h}}$
  theoretical and experimental transition energies in eV. The
  superscripts stand for: a-Ref.~\cite{897}, b-Ref.~\cite{913},
  c-Ref.~\cite{898}, d-Ref.~\cite{784}, e-Ref.~\cite{900,899},
  f-Ref.~\cite{890,1000} and g-Ref.~\cite{788}. Since the
  K$\alpha_{1,2}^{\text{h}}$ theoretical values obtained by Chen et
  al. \cite{308} were presented in the form of energy shifts from the
  corresponding diagram lines, we used the K$\alpha_{1,2}$ energy
  values calculated by the same authors \cite{829} in order to obtain the
  K$\alpha_{1}^{\text{h}}$ and the K$\alpha_{2}^{\text{h}}$ transition
  energies. An asterisk indicates that the authors presented
  experimental results asenergy shifts from the corresponding diagram
  line. In these cases we took the diagram line energy values
  from~\cite{3000}.  }
\label{tab:all_z_kah}                                              
\scriptsize 

\begin{tabular}{cccccccccccc}
\hline
\hline
&  &  &  &  &  &  &  &  &  \\ 
& \multicolumn{6}{c}{K$\alpha _{2}^{\text{h}}$} &  &
\multicolumn{4}{c}{K$\alpha_{1}^{\text{h}}$} \\ 
\cline{2-7}\cline{9-12}
& \multicolumn{5}{c}{Theory} & Experiment &  & \multicolumn{3}{c}{Theory} & 
Experiment \\ 
\cline{2-6}\cline{9-11}
$Z$ & MCDF & Fitted & Ref.~\cite{896} & Ref.~\cite{880} &
Ref.~\cite{308} &  &  & MCDF & Fitted & Ref.~\cite{308} \\ 
\hline
&  &  &  &  &  &  &  &  &  \\ 
12 & 1368.53 &      & 1381 &      &        & 1367.8$\pm $0.2$^{\text{a}}$ &  & 1374.34 &  &  \\ 
13 & 1611.75 &      & 1627 & 1608 &        &  $\ $1610.8$\pm $0.2$^{\text{b*}}$ &  & 1616.69  &  &  \\ 
14 &         & 1874 & 1893 &      &        &  &  &   & 1880 &  \\ 
15 &         & 2157 & 2179 &      &        &  &  &   & 2164 &  \\ 
16 &         & 2461 & 2487 &      &        &  &   &  & 2469 &  \\ 
17 &         & 2785 & 2816 & 2777 &        &  &  &  & 2794 &  \\ 
18 & 3131.50 & 3164 &      &      & 3130.9 &   &  & 3141.62 &  & 3141.7 \\ 
19 &         & 3498 &      &      &        &  &   & 3508 & &  \\ 
20 & 3884.80 &      &      & 3864 & 3884.4 &   &  & 3896.39 &  & 3896.8 \\ 
21 & 4294.16 &      &      &      &        &  &   & 4306.27 & &  \\ 
22 & 4723.86 &      &      &      &        & 4727$\pm $2$^{\text{c}}$ &  & 4736.76 &  &   & 4741$\pm $3$^{\text{c}}$  \\ 
23 &         & 5174 &      &      &        &  &    & & 5188 &  \\ 
24 &         & 5647 &      &      &        & 5650$\pm $2$^{\text{c}}$ &  &  & 5662 &   & 5666$\pm $3$^{\text{c}}$ \\ 
   &         &      &      &      &        & 5645$\pm $2$^{\text{d}}$ &  &  &      &  \\ 
25 &         & 6140 &      &      & 6140.1 & 6160$\pm $20$^{\text{e}}$ &  &  & 6156 & 6158.0 \\ 
26 &         & 6655 &      & 6597 &        & 6659$\pm 2^{\text{c}}$ &  &  & 6673 &   & 6679$\pm $3$^{\text{c}}$ \\ 
   &         &      &      &      &        & 6655$\pm 2^{\text{d}}$ &  &  &      &   & 6675$\pm $2$^{\text{d}}$ \\ 
   &         &      &      &      &        & 6658$\pm 2^{\text{e*}}$ &  &  &      &   & 6677.36$\pm $0.18$^{\text{f}}$ \\ 
   &         &      &      &      &        & 6658.31$\pm 0.10^{\text{f}}$ &  &  &      &  \\ 
27 &         & 7191 &      &      &        & 7192$\pm 3^{\text{d}}$ &  &  & 7211 &   & 7207$\pm $3$^{\text{d}}$ \\ 
28 & 7749.04 &      &      & 7670 &        & 7751$\pm 2^{\text{c}}$ &  & 7770.90 &  &   & 7775$\pm $3$^{\text{c}}$ \\ 
   &         &      &      &      &        & 7752$\pm $3$^{\text{e*}}$ &  &  &  &  \\

29 &         & 8328 &      &      &        & 8331$\pm 3^{\text{d}}$ &  &  & 8352 &   & 8352$\pm $3$^{\text{d}}$ \\ 
   &         &      &      &      &        & 8331$\pm $3$^{\text{e*}}$ & & &  &  & 8353.1$\pm $0.7$^{\text{f}}$   \\ 
   &         &      &      &      &        & 8329.5$\pm $0.3$^{\text{f*}}$ &  &  &  &  \\ 
   &         &      &      &      &        & 8362$\pm 17^{\text{g}}$ &  &  &  &  \\

30 & 8928.53 &      &      &      & 8928.7 &         & & 8954.97 &  & 8955.4 \\ 
&  &         &      &      &      &        &  &  &  \\ 
\hline\hline
\end{tabular}                                       
\end{center}                                     
\end{table}                                      




\begin{table}                                               
\begin{center}                                              
\caption{K$\alpha _{2}\alpha _{3}$ and K$\alpha _{1}\alpha _{3}$
  theoretical and experimental transition energies in eV. The
  superscripts stand for: a-Ref.~\cite{783}, b-Ref.~\cite{901,784} and
  c-Ref.~\cite{786}.  }
\label{tab:all_z_kaa}                                              
\scriptsize 

\begin{tabular}{cccccccccc}
\hline\hline
&  &  &  &  &  &  &  &  &  \\ 
& \multicolumn{5}{c}{K$\alpha _{2}\alpha _{3}$} &  & \multicolumn{3}{c}{K$\alpha _{1}\alpha _{3}$} 
\\ \cline{2-6}\cline{8-10}
& \multicolumn{4}{c}{Theory}  & Experiment &  & \multicolumn{2}{c}{Theory}  & Experiment \\ 
\cline{2-5}\cline{8-9}
$Z$ & MCDF & Fitted & Ref.~\cite{896} & Ref.~\cite{880} &  &  & MCDF  & Fitted &  \\ \hline
&  &  &  &  &  &  &  &  &  \\ 
12 & 2585.45 &  & 2653 &  &  &  & 2600.81 &  &  \\ 
13 & 3056.54 &  & 3188 & 3051 & 3076.8$^{\text{a}}$ &  & 3072.10 &  &  \\ 
14 &  & 3566 & 3673 &  &  &  &  & 3584 &  \\ 
15 &  & 4118 & 4240 &  &  &  &  & 4137 &  \\ 
16 &  & 4710 & 4848 &  &  &  &  & 4731 &  \\ 
17 &  & 5345 & 5489 & 5333 &  &  &  & 5367 &  \\ 
18 & 6022.29 &  & 6178 &  &  &  & 6046.71 &  &  \\ 
19 &  & 6378 &  &  &  &  &  & 6764 &  \\ 
20 & 7497.79 &  &  & 7464 &  &  & 7525.09 &  &  \\ 
21 & 8300.74 &  &  &  &  &  & 8328.97 &  &  \\ 
22 & 9145.04 &  &  &  &  &  & 9174.72 &  &  \\ 
23 &  & 10029 &  &  &  &  &  & 10060 &  \\ 
24 &  & 10958 &  &  & 10935$\pm $8$^{\text{b}}$ &  &  & 10990 & 10960$\pm $8$%
^{\text{b}}$ \\ 
25 &  & 11929 &  &  & 11907$\pm $20$^{\text{c}}$ &  &  & 11962  &  \\ 
26 &  & 12942 &  & 12483 & 12907$\pm 9^{\text{b}}$ &  &  & 12978 & 12953$\pm 
$9$^{\text{b}}$ \\ 
27 &  & 13999 &  &  & 13945$\pm 10^{\text{b}}$ &  &  & 14036 & 14005$\pm $10$%
^{\text{b}}$ \\ 
28 & 15098.34 &  &  & 14960 & 15060$\pm 10^{\text{b}}$ &  & 15136.33 &  & 
15108$\pm $10$^{\text{b}}$ \\ 
29 &  & 16240 &  &  & 16193$\pm 10^{\text{b}}$ &  &  & 16281 & 16236$\pm $10$%
^{\text{b}}$ \\ 
30 & 17423.97 &  &  &  &  &  & 17466.82 &  &  \\ 
&  &  &  &  &  &  &  &  &  \\ 
\hline\hline
\end{tabular}                                       
\end{center}                                     
\end{table}                                      


%
%
\ 
\newpage
%

%
%

%
%

%
%
%
%

%
%
\end{document}